\documentclass[aps,prd,reprint,superscriptaddress,nofootinbib]{revtex4-1}

\usepackage{amsmath}
\usepackage[dvipsnames]{xcolor}
\usepackage{graphicx}
\usepackage{hyperref}
\usepackage{mathtools}
\usepackage{comment}
\usepackage{cancel}
\usepackage{physics}
\usepackage{mathtools}
\usepackage{adjustbox}
\usepackage{multirow,array,booktabs}
\usepackage[shortlabels]{enumitem}
\usepackage{enumitem} 

\hypersetup{
 bookmarksnumbered=true,
 colorlinks=true,
 linkcolor=[rgb]{0.098,0.098,0.439},
 citecolor=[rgb]{0.098,0.098,0.439},
 urlcolor=[rgb]{0.098,0.098,0.439}
  }

\begin{document}

\title{Wave-envelope dark matter beyond the monochromatic paradigm}
\author{Yechan Kim}
\email{cj7801@kaist.ac.kr}
\affiliation{Department of Physics, Korea Advanced Institute of Science and Technology, Daejeon 34141, Korea}
\author{Hye-Sung Lee}
\email{hyesung.lee@kaist.ac.kr}
\affiliation{Department of Physics, Korea Advanced Institute of Science and Technology, Daejeon 34141, Korea}
\date{April 2026}

\begin{abstract} 
Ultralight dark matter searches widely assume that signals are monochromatic, with a single frequency set by the mass.
This assumption is generally violated in the presence of field mixing, even when the constituent fields have similar frequencies. 
Instead, dark matter signals can exhibit a two-timescale structure with intrinsic slow modulation.
We demonstrate that mixing between ultralight wave dark matter fields induces a parametric structure, leading to a scenario we refer to as wave-envelope dark matter, in which a slow-beating envelope emerges alongside the primary oscillation. 
This results in distinctive features such as slow modulation and characteristic sideband structures in the frequency spectrum, beyond the conventional monochromatic expectation. 
As a representative example, we briefly discuss implications for neutrino observables.
\end{abstract}

\maketitle

\section{Introduction} 

Ultralight dark matter can behave as a coherently oscillating classical field, commonly referred to as wave dark matter (wave DM)~\cite{Hu:2000ke, Hui:2021tkt}. 
In many settings, such oscillations induce time-dependent signals through couplings to Standard Model fermions~\cite{Arvanitaki:2016fyj, Kennedy:2020bac, Kobayashi:2022vsf, Filzinger:2023qqh, Kim:2025vnc}, gauge bosons~\cite{VanTilburg:2015oza, Guo:2022vxr, Flambaum:2024zyt}, or sterile neutrinos~\cite{Berlin:2016woy, Krnjaic:2017zlz, Losada:2022uvr, Dev:2022bae, Huang:2021kam, ChoeJo:2023ffp, Martinez-Mirave:2024dmw, Kim:2025xum}. 
A central assumption underlying these searches is that the signal is monochromatic, with a single frequency set by the wave DM mass.

In this work, we point out that this assumption is generally violated when the DM consists of mixed wave fields, even when the constituent fields have similar oscillation frequencies.
Instead, the DM background can exhibit a two-timescale structure, where a slow intrinsic modulation accompanies the primary oscillation, leading to characteristic sideband structures in the frequency spectrum.
We propose a scenario, \textit{wave-envelope dark matter}, in which a slow-beating envelope emerges together with the primary oscillation.
This behavior should be distinguished from ordinary beating, as the modulation arises from parametric dynamics induced by field mixing rather than from the superposition of independent oscillatory modes.

The mixing of wave fields induces time-dependent effective masses and leads to a parametric structure.
The parametric structure can be described by a Mathieu-type equation, similar to preheating after inflation~\cite{Kofman:1994rk, Kofman:1997yn, Greene:1997fu}. 
Depending on the parameters, the system can enter a narrow resonance regime or a stable region, where parametric amplification is inefficient or absent.
In this regime, the system exhibits a slow-beating envelope behavior rather than exponential amplification. 
This defines a distinct dynamical regime, in which the parametric response leads to coherent amplitude modulation instead of exponential growth.  

This additional slow modulation modifies the conventional single-frequency interpretation of DM signals. 
If wave-envelope DM interacts with other particles, the induced signals inherit this behavior. 
As a representative example, we briefly discuss implications for the neutrino sector.

The parametric resonance of ultralight wave DM has been examined for axion resonance~\cite{Greene:1998pb, Soda:2017dsu, Fukunaga:2019unq, Co:2020dya, Nakayama:2021avl}, vector field production~\cite{Dror:2018pdh, Masaki:2019ggg, Carenza:2019vzg, Arza:2020eik}, and gravitational wave amplification~\cite{Kitajima:2018zco, Delgado:2023psl}.
The slow-beating behavior may appear in preheating in the presence of non-canonical kinetic terms~\cite{Underwood:2013pwa}. 
The mixing effects of ultralight wave DM have also been studied in the context of multi-axion models~\cite{Hill:1988bu, Kim:2004rp, Daido:2015cba}, solitonic cores at stellar/galactic scales~\cite{Guo:2020tla, Glennon:2023jsp, Mirasola:2024pmw}, and black hole superradiance~\cite{Fukuda:2019ewf, Lyu:2025lue, Zhu:2025enp}.
In our work, we focus on slow envelope modulation of ultralight wave DM arising from field mixing, rather than on resonance-driven dynamics.

\section{Mixing of ultralight wave dark matter}
The wave DM has an ultralight mass scale in the range of $3 \times 10^{-21}\,\mathrm{eV}$ to $30\,\mathrm{eV}$, where the lower bound originates from Lyman-$\alpha$ forest and the upper bound comes from large occupation number and longer de Broglie wavelength than inter-particle separation requirements to treat as a classical wave~\cite{Hu:2000ke, Hui:2021tkt}.
In the expanding universe under the assumption of spatial homogeneity, the equation of motion for a real scalar wave DM field $\phi$ is
\begin{align}
\ddot{\phi} + 3 H \dot{\phi} + m_\phi^2 \phi & = 0 ,
\label{EoM0}
\end{align}
where $H$ is the Hubble parameter and $m_\phi$ is the wave DM mass. 
In the late universe, the Hubble friction is negligible compared to the DM mass. 
The wave DM $\phi$ satisfies the following harmonic oscillating solution.
\begin{align}
\phi(t) & = \frac{\sqrt{2 \rho_\phi}}{m_\phi} \, \sin (m_\phi t),
\label{Sol0}
\end{align}
where $\rho_\phi$ is the energy density of the wave DM. The energy density reduces as $\rho_\phi \propto a^{-3}$ over cosmic history, where $a$ is the scale factor.
Thus, the oscillation amplitude of the wave DM also decreases.
The oscillation of the wave DM can be transferred to the other particles via interactions, modulating their masses or couplings~\cite{VanTilburg:2015oza, Arvanitaki:2016fyj, Berlin:2016woy, Krnjaic:2017zlz, Kennedy:2020bac, Huang:2021kam, Dev:2022bae, Losada:2022uvr, Guo:2022vxr, Kobayashi:2022vsf, ChoeJo:2023ffp, Filzinger:2023qqh, Martinez-Mirave:2024dmw, Flambaum:2024zyt, Kim:2025xum, Kim:2025vnc}.

We consider mixing between two wave DM scalar fields, $\phi$ and $\Phi$.
\begin{align}
V(\phi, \Phi) = \frac{1}{2} m^2 \phi^2 
+ \frac{1}{2} M^2 \Phi^2
+ \frac{1}{2} \kappa \phi^2 \Phi^2 ,
\label{Lagrangian}
\end{align}
where $m$ and $M$ are the bare masses of the wave DM $\phi$ and $\Phi$, respectively, and $\kappa$ is the coupling between them. 
The coupled equations of motion for $\phi$ and $\Phi$ are
\begin{align}
\ddot{\phi} + 3 H \dot{\phi} + (m^2 + \kappa \Phi^2) \phi & = 0 ,
\label{EoM1}
\\
\ddot{\Phi} + 3 H \dot{\Phi} + (M^2 + \kappa \phi^2) \Phi & = 0 .
\label{EoM2}
\end{align}
The mixing term enables energy transfer between $\phi$ and $\Phi$.
Under the mixing interaction, the effective masses of the wave DM fields become time-dependent.
\begin{align}
m_\phi(t) & = \sqrt{m^2 + \kappa \Phi^2(t)}, \;
\label{mphi}
\\
M_\Phi(t) & = \sqrt{M^2 + \kappa \phi^2(t)} .
\label{MPhi}
\end{align}

The mixing term should be subdominant to the bare mass terms to avoid radiation-like behavior~\cite{Turner:1983he, Das:2020nwc, Cembranos:2018ulm}.  
In the early universe, the large amplitude of wave DM can render the mixing term dominant, potentially modifying the standard matter-like cosmological evolution.
This can be avoided if the mixing interaction becomes relevant only at late times.
As an illustrative realization, consider the potential $V\supset \frac12 S^2 \phi^2 \Phi^2 / \Lambda^2$, where $\Lambda$ is the cutoff.
If the scalar $S$ acquires a vacuum expectation value only at late times, the effective coupling is given by $\kappa = \langle S \rangle^2 / \Lambda^2$.
Such a scenario can arise from a temperature-dependent potential or a Hubble-induced mass term for $S$.
In this way, the mixing interaction remains negligible during the early cosmological evolution and becomes relevant only in the late universe. We do not specify or analyze the detailed UV model here.

\section{Mathieu analysis and parametric structure}

\begin{figure}[tb]
    \centering
    \includegraphics[width=0.45\textwidth]{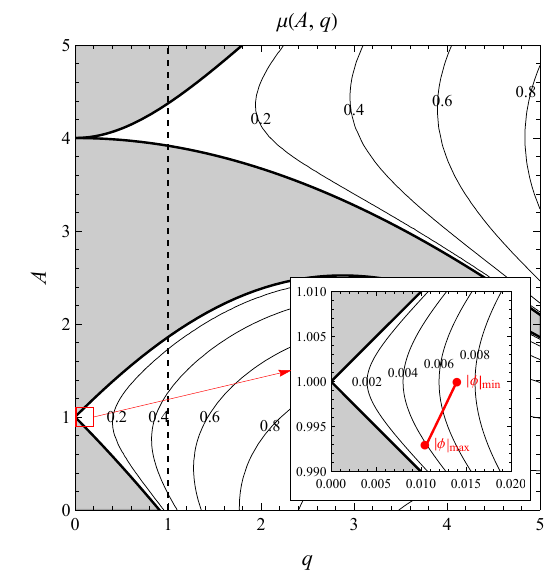}
    \caption{
    The instability bands (white regions) in the $(A,q)$ plane, together with contours of the Mathieu characteristic exponent $\mu(A,q)$. 
    The inset highlights the narrow resonance regime ($q \ll 1$) around $A \simeq 1$, near the boundary of the instability band.  
    The red line shows the trajectory corresponding to the representative benchmark in Fig.~\ref{phiPlot}.
    }
    \label{AqPlot}
\end{figure}

The dynamics of the mixed system can be described by a Mathieu-type equation, exhibiting parametric structure depending on the parameters.
For a static mass $M_\Phi$ in the limit of $M^2 \gg \kappa \phi^2$, the wave DM $\Phi$ follows the harmonic oscillating behavior $\Phi(t) \simeq \Phi_0 \sin (M t)$ approximately.
Introducing the variable $z \equiv M t$ yields the linear Mathieu equation for $\phi$~\cite{Kofman:1994rk}
\begin{align}
\phi''(z) + \big[ A - 2q \cos (2z) \big] \phi(z) = 0,
\label{eq:MathieuEqn}
\end{align}
where $' = d/dz$, and
\begin{align}
q \equiv \frac{\kappa \Phi_0^2}{4 M^2}, \quad A \equiv \frac{m^2}{M^2} + 2q.
\label{qAparameter}
\end{align}
The solution can be written as
\begin{align}
\phi(z) = e^{\tilde \mu(A,q) \, z} \, P(z),  
\end{align}
where $\tilde \mu(A,q) = \mu + i \nu$ is the Mathieu characteristic and $P(z)$ is a periodic function with period $\pi$.
The instability bands are characterized by $\mu>0$, for which the field grows exponentially, $|\phi(z)| \propto e^{\mu  z }$ during the linear resonance regime.
In the following, we refer to $\mu$ as the Mathieu characteristic exponent.

Figure~\ref{AqPlot} describes the instability bands of the Mathieu equation over the parameter $(A, q)$.
In the broad resonance with $q \gg 1$, the instability bands are wide.
$\mu$ is also large, which leads to efficient amplification.
The broad resonance regime is typically adopted for the preheating after inflation~\cite{Kofman:1994rk, Kofman:1997yn, Greene:1997fu} or amplified productions of particles~\cite{Greene:1998pb, Soda:2017dsu, Fukunaga:2019unq, Co:2020dya, Nakayama:2021avl, Dror:2018pdh, Masaki:2019ggg, Carenza:2019vzg, Arza:2020eik, Kitajima:2018zco, Delgado:2023psl}.

In contrast, for the narrow resonance regime with $q \ll 1$, the instability regions are thin and centered around $A \simeq n^2$ ($n=1,2, \cdots$). 
For instance, the first band around $n=1$ is approximated as $|A-1| \lesssim q$, and the Mathieu characteristic exponent becomes $\mu \simeq \frac12 \sqrt{q^2 - (A-1)^2}$.
For higher $n$-th band, exponent is more suppressed as $\mu \propto q^n$.
For the narrow resonance, $\mu$ is small, so the amplification is weaker. 
Close to the boundary of the instability band, the growth rate is suppressed, and the dynamics deviates from the conventional resonance phase characterized by exponential amplification.

Within the instability band, the imaginary part of the Mathieu characteristic is fixed as $\nu = n$ for the $n$-th band, which does not introduce an additional dynamical modulation. 
In the stable region, $\mu =0$ and $\nu$ deviates from $n$, and this small deviation governs dynamics.
For example, near the $n=1$ band, the imaginary part becomes $\nu \simeq 1 \pm \frac{1}{2} \sqrt{(A-1)^2 - q^2}$, where the sign depends on whether $A$ is larger than 1 or not.

\begin{figure}[tb]
    \centering
    \includegraphics[width=0.48\textwidth]{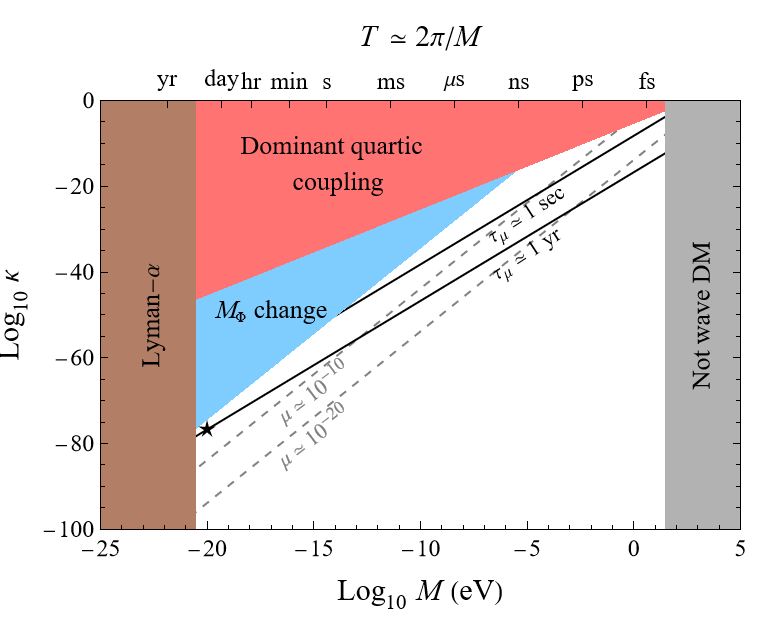}
    \caption{
    Parameter space of the wave DM mass $M$ and the mixing coupling $\kappa$. 
    The oscillation period of the wave DM is given by $T \simeq 2\pi/M$ (top axis). 
    Gray dashed lines show contours of the Mathieu characteristic exponent $\mu$, while black lines indicate the slow modulation timescale $\tau_\mu \simeq 2\pi/(\mu M)$. 
    The star marks the representative benchmark used in Fig.~\ref{phiPlot}.
    }
    \label{ParamSpace}
\end{figure}

In our scenario, the mixing term is subdominant to the bare mass terms, so the system can naturally lie in the narrow resonance regime with $q \ll 1$. 
We focus on the first band with $A \simeq 1$ as a representative example.
The complementary stable-region realization is discussed separately in Sec.~\ref{sec:stable}.
In the first band, two wave DM fields naturally get similar bare masses $m\simeq M$.
In general, the following discussion can be extended to other narrow resonance bands, but the higher $n$-th band is much narrower and suppressed.

Figure~\ref{ParamSpace} shows the parameter space of the wave DM mass $M$ and the mixing coupling $\kappa$.
We fix the density ratio to $\rho_\Phi = 10\,\rho_\phi$, and choose $m$ such that $A=1$ at the initial time.
The combined contribution of the two wave DM components accounts for the local DM density $\rho_\mathrm{DM} = 0.4\,\mathrm{GeV/cm}^3$~\cite{Hui:2021tkt}.
The brown and gray regions indicate the lower and upper bounds on the wave DM mass, respectively.
The red region corresponds to the constraint that the radiatively generated quartic coupling remains subdominant~\cite{Turner:1983he, Das:2020nwc}.
In the blue region, the effective mass $M_\Phi$ becomes time-dependent due to mixing.

Although the value of the mixing coupling $\kappa$ appears extremely small, its scale should be considered relative to the large amplitude of ultralight wave DM.
We define a reference coupling by requiring the mixing-induced contribution to the mass squared of $\phi$ to equal its bare mass term, $\kappa_{\rm ref} = m^2/\Phi_0^2 = m^2 M^2 / (2 \rho_\Phi)$.
For $\kappa \ll \kappa_{\rm ref}$, the mixing effect is too small to generate distinct dynamics.
In contrast, for $\kappa \gtrsim \kappa_{\rm ref}$, the mixing contribution can be larger than the bare term, which is outside the regime we considered here.  
Therefore, the phenomenologically relevant behavior is realized within a finite range below $\kappa_{\rm ref}$.
A similar situation arises for quartic self-interactions of ultralight wave DM, where a finite window of extremely small couplings can still be dynamically relevant owing to the large field amplitude~\cite{Cembranos:2018ulm, Garcia:2023abs, Koo:2025jkx}.

\section{Wave-envelope dynamics in instability bands}

In the narrow-resonance regime, sustained exponential amplification is suppressed by the small growth rate and nonlinear backreaction. 
Instead, the system develops a slow
envelope modulation of the oscillation amplitude.
This behavior is characterized by two distinct timescales.
The oscillating period $T \simeq 2\pi / M$ of the wave DM is determined by its mass.
On the other hand, the solution of the Mathieu equation contains a factor $e^{\mu z}$ with $z = Mt$, which introduces a slower timescale associated with $\mu M$.
A new timescale $\tau_\mu$ is associated with the Mathieu characteristic exponent $\mu$.
For $\mu \ll 1$, this leads to a parametrically separated timescale
\begin{align}
T \simeq \frac{2\pi}{M} \ll \tau_\mu \simeq \frac{2\pi}{\mu M}.
\end{align} 
The longer timescale $\tau_\mu$ controls the envelope modulation of the wave DM amplitude.

\begin{figure}[tb]
    \centering
    \includegraphics[width=0.48\textwidth]{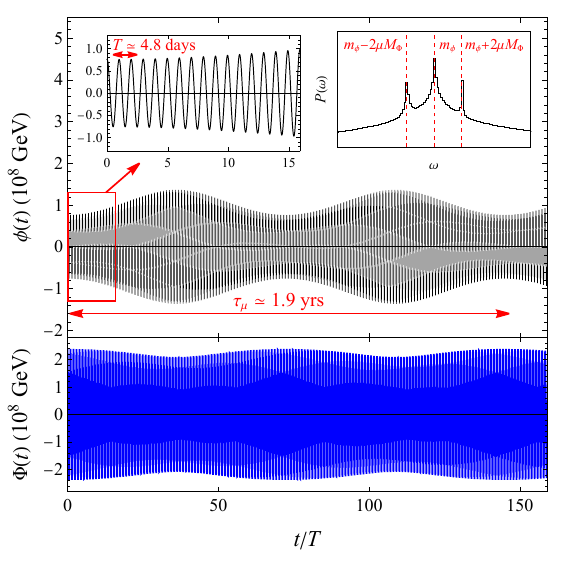}
    \caption{
    Time evolution of wave-envelope DM $\phi(t)$ exhibiting two distinct timescales. 
    The main upper panel shows the slow envelope modulation with timescale $\tau_\mu \simeq 2\pi/(\mu M)$, while the inset displays the fast oscillation with period $T \simeq 2\pi/M$. 
    An additional inset shows the frequency spectrum $P(\omega)$, where the intrinsic slow modulation generates characteristic sidebands at $m_\phi \pm 2\mu M_\Phi$.
    The amplitude of $\Phi(t)$ in the lower panel also modulates slightly due to the backreaction.
    }
    \label{phiPlot}
\end{figure}

Figure~\ref{phiPlot} shows the behavior of $\phi(t)$ as an illustrative benchmark in the narrow resonance regime. 
The coupled equations in Eqs.~\eqref{EoM1}--\eqref{EoM2} are solved for a representative choice of parameters: $M = 10^{-20}\,\mathrm{eV}$, $\kappa = 10^{-76}$ ($\simeq 0.06\, \kappa_{\rm ref}$), and $m$ chosen such that $A = 1$, with $\rho_\Phi = 10\,\rho_\phi$ at the initial time.
The total energy density is fixed to match the local DM density, and the initial kinetic energies are set to zero.
The particular value of $\kappa$ is chosen to illustrate a distinct longer timescale relevant for long-duration observations, rather than being predicted by the effective theory.
$\kappa$ can receive a self-induced one-loop correction, but it is negligible for our benchmark since the correction is proportional to $\kappa^2$.
No additive correction arises because there is no additional field coupled to both components.

The parameter setup yields $\mu \simeq 0.007$ at the initial time.
The fast oscillation period from the wave DM mass is $T \simeq 2\pi / M \simeq 4.8\,\mathrm{days}$, and the slow oscillation appears with $\tau_\mu \simeq 2 \pi / (\mu M) \simeq 1.9\,\mathrm{yrs}$.
Note that $m$ is of the same order as $M$, corresponding to the narrow resonance condition $|A-1| \lesssim q$, which implies a near-degeneracy $|m - M| \lesssim \mathcal{O}(q M)$.

The amplitude of $\Phi(t)$ also varies due to the backreaction from $\phi(t)$, inducing a time dependence in $A(t)$ and $q(t)$, as shown in Fig.~\ref{AqPlot}. 
Although the system deviates from an exact Mathieu equation due to the time dependence of the parameters, its local behavior remains well approximated by an instantaneous Mathieu form, which captures the relevant dynamical timescales.
Near the boundary of the instability band, the Mathieu characteristic exponent $\mu$ is small, making the parametric amplification inefficient.
Together with the time-dependent parameters, this prevents sustained exponential growth.
Consequently, the amplitude of $\phi$ develops a slow envelope modulation rather than exponential amplification.
This defines a distinct dynamical regime arising from the parametric structure rather than ordinary beating, which we refer to as \textit{wave-envelope dark matter}.

The characteristic two-timescale structure of the wave-envelope DM can be captured by the following leading-order expression
\begin{align}
\phi(t) & \simeq \phi_0(t) \,\sin (m_\phi t),
\\
\phi_0(t) & \simeq  
\frac{|\phi|_\mathrm{max} + |\phi|_\mathrm{min}}{2} + \frac{|\phi|_\mathrm{max} - |\phi|_\mathrm{min}}{2} \sin(2\mu M_\Phi \,t + \theta),
\label{phi0t}
\end{align}
for $A \simeq 1$ and $q\ll 1$.
Here, $|\phi|_\mathrm{max}$ and $|\phi|_\mathrm{min}$ are the maximum and minimum amplitudes of $\phi$, respectively, and $\theta$ is the phase offset.
Small $\mu$ in the narrow resonance regime can generate a well-separated long timescale $\tau_\mu \gg T$, whereas the fractional amplitude modulation is relatively modest.
The fast oscillation is set by the wave DM mass $m_\phi$, while the slow oscillation frequency is determined by $2\mu M_\Phi$, where a factor of 2 follows from the cycle-counting convention for $\tau_\mu$, in which one full envelope cycle includes two
successive modulations.
Small deviations from this leading-order expression can arise due to the time-dependent effective wave DM masses.

As shown in Fig.~\ref{phiPlot}, this two-timescale structure directly translates into a characteristic frequency spectrum at $m_\phi$ and sidebands $m_\phi \pm 2\mu M_\Phi$, originating from the intrinsic slow modulation of the amplitude.
Since the fractional amplitude modulation is modest, the power carried by the sidebands might be correspondingly suppressed.
Nevertheless, their symmetric placement and equal frequency spacing provide a distinctive spectral signature, encoding the mixing-induced parametric dynamics.

The wave DM generically exhibits stochastic amplitude and phase variation with the coherence time $t_{\rm coh} \simeq 2\pi/(m_\phi v^2_{\rm DM})$ due to its velocity dispersion~\cite{Hui:2021tkt}.
For $v_{\rm DM} \sim 10^{-3}$, the coherence time is significantly longer than the envelope oscillation time $t_{\rm coh} \gg \tau_\mu \gg T$ in our benchmark.
The generic modulation is stochastic because it originates from the superposition of modes with a velocity distribution.
By contrast, for the near-band benchmark considered here, the envelope from the mixing interaction remains coherent and periodic over cycles.
It can be distinguished through its regular temporal structure and the associated sideband frequencies, rather than through an irregular drift.

Although $\phi$ constitutes a relatively subdominant DM component in our setup, its initial amplitude is not strongly suppressed. 
For $\rho_\Phi = 10 \rho_\phi$ and $m\simeq M$, the initial amplitudes satisfy $\phi_0/\Phi_0 \simeq \sqrt{\rho_\phi/\rho_\Phi} \simeq 0.3$.
The fraction of $\phi$ varies throughout the envelope evolution due to the energy exchange with $\Phi$.
In addition, the initial abundance of $\phi$ alone does not determine the observability of the modulation signal;
its absolute signal amplitude also depends on the coupling of $\phi$ to the probed sector.

\section{Wave-envelope dynamics in stable regions}
\label{sec:stable}

\begin{figure}[tb]
    \centering
    \includegraphics[width=0.45\textwidth]{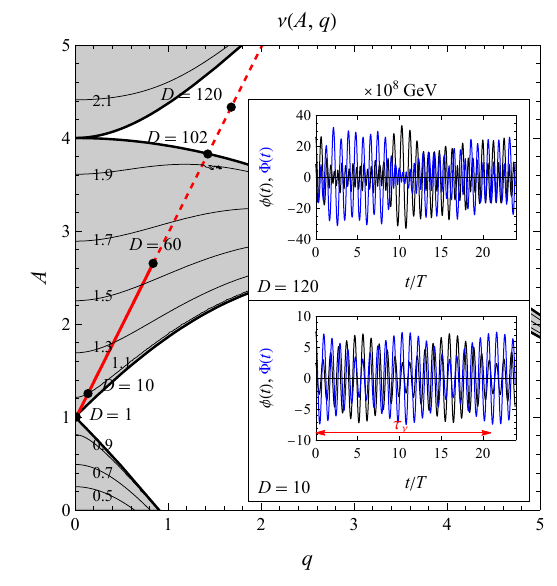}
    \caption{
    The imaginary part $\nu(A,q)$ of the Mathieu characteristic inside the stable regions (gray regions). 
    The red line represents the trajectory of $(A,q)$ along the enhancement factor $D$ starting from the benchmark in Fig.~\ref{phiPlot}. 
    For the stable region near the $n$-th band, $\nu$ deviates from $n$, and the small deviation determines a new longer timescale $\tau_\nu$, as shown in the inset for $D=10$.
    For higher enhancement $D\gtrsim 60$, the Mathieu approximation becomes less valid, as illustrated by the irregular evolution in the $D=120$ inset.
    The density-enhancement trajectory is discussed further in Appendix~\ref{sec:dense}.
    }
    \label{AqPlotDen}
\end{figure}

For the unstable region, the slow oscillating envelope behavior in the narrow resonance is governed by the real part $\mu \ll 1$ of the Mathieu characteristic, and the imaginary part $\nu = n$ for the $n$-th band is irrelevant.
In contrast, in the stable region, the roles of the real and imaginary parts are reversed; $\mu = 0$, but $\nu$ deviates from $n$ near the $n$-th band, providing a new modulation timescale.
Figure~\ref{AqPlotDen} shows the imaginary part $\nu$ inside the stable regions over the Mathieu parameter space $(A,q)$.

In the stable region with vanishing real part $\mu=0$, the solution of the Mathieu-type equation can be expanded as~\cite{Fruchting:1969, kovacic2018mathieu, verhulst2022perturbation}
\begin{align}
\phi = e^{i \nu z} \sum_{l=-\infty}^\infty c_l e^{2 i l z} ,
\end{align}
since $P(z)$ has period $\pi$.
Inserting this expansion into Eq.~\eqref{eq:MathieuEqn} gives
\begin{align}
-(\nu + 2l)^2 c_l + A c_l - q (c_{l-1} + c_{l+1} ) = 0 .
\end{align}
Near the $n$-th band with $(A,q) \simeq (n^2,0)$, the dominant modes are $l=0$ and $l=-n$, corresponding to the frequencies $\nu$ and $\nu-2n$, respectively.
Since $\nu = n+\delta$ deviates from $n$ with small $|\delta| \ll 1$, the real solution is given by an effective two-mode description with frequencies $n+\delta$ and $n-\delta$.
The interference between these two modes introduces a new slow modulation timescale $\tau_\nu$ compared to the original wave DM period $T$,
\begin{align}
T \simeq \frac{2\pi}{M} \ll
\tau_\nu \simeq \frac{4\pi}{|\nu-n| M},
\end{align}
arising from the small deviation $|\delta| = |\nu-n| \ll 1$ in the stable region adjacent to the $n$-th band.
Unlike ordinary beating between independent free fields, the two nearby frequencies here are generated by the mixing-induced parametric structure.
An additional factor of 2 arises from adopting the same convention as for $\tau_\mu$, where one envelope cycle comprises two oscillations of the amplitude.
When the deviation becomes large away from the band boundary, the timescale is not parametrically separated from the original oscillation period.

Wave-envelope behavior can arise not only from a suppressed exponent in the instability band, as discussed in the previous section, but also from adjacent stable regions.
The linear Mathieu analysis provides a parametrically longer timescale in both regimes, controlled by $\mu$ inside the instability band and by $|\nu-n|$ in the neighboring stable region.

The stable-region analysis above can be directly applied when the DM density increases from our benchmark with enhancement factor $D$ for the initial amplitudes, as discussed in detail in Appendix~\ref{sec:dense}. 
However, the stable-region realization is not specific to density-enhanced environments; it can also arise directly for other parameter setups.

Interactions with oscillating wave DM can induce modulation signals in a wide range of experimental probes, including atomic clocks~\cite{Arvanitaki:2016fyj, Kennedy:2020bac, Kobayashi:2022vsf, Filzinger:2023qqh}, precision spectroscopy~\cite{VanTilburg:2015oza}, and neutrino observables~\cite{Berlin:2016woy, Krnjaic:2017zlz, Losada:2022uvr, Dev:2022bae, Huang:2021kam, ChoeJo:2023ffp, Martinez-Mirave:2024dmw, Kim:2025xum}. 
These searches typically assume that the DM background oscillates with a single frequency set by its mass. 
In contrast, for wave-envelope DM, the induced signal follows $\phi(t)$ and inherits the intrinsic slow modulation, leading to characteristic sideband structures in the frequency spectrum. 
This provides a distinctive signature beyond the conventional monochromatic expectation.

As a representative example, one can consider couplings to neutrinos, where the wave-envelope DM field generates a time-dependent Majorana mass. 
This can induce periodic transitions between Dirac and Majorana types, depending on the relative size of the Dirac and Majorana mass terms~\cite{ChoeJo:2023ffp}. 
In particular, when the system enters the Dirac type, lepton-number-violating processes are suppressed~\cite{Doi:1982dn, Schechter:1981bd, Bolton:2019pcu}, leading to a time-dependent modulation of observables such as neutrinoless double beta decay. 
The duration and timing of these suppression intervals are controlled by the slow envelope, reflecting the intrinsic two-timescale structure of the underlying DM dynamics. 
Further details are provided in Appendix~\ref{appenNu}.
This exemplifies how the intrinsic modulation of wave-envelope DM can lead to qualitatively new time-dependent signatures in experimental observables.

Axion provides another application of wave-envelope DM.
Conventional axion searches typically target a single narrow spectral feature with a monochromatic frequency at each trial mass in haloscope or spin-precession experiments~\cite{Sikivie:1983ip, Budker:2013hfa, Ouellet:2018beu, ADMX:2020ote, CAPP:2024dtx}.
For multi-axion models~\cite{Hill:1988bu, Kim:2004rp, Daido:2015cba}, the mixing interaction can arise from the small-field expansion of coupled cosine potentials.
If the axion exhibits wave-envelope dynamics through its mixing, the induced signal inherits the slow amplitude modulation and develops correlated sidebands around the center frequency.
Accordingly, axion searches could be extended from a single-line template to a multi-frequency analysis targeting a symmetrically spaced set of peaks, encoding the additional envelope timescale.

\section{Summary and outlook}
Ultralight wave DM is typically assumed to generate monochromatic signals with a single frequency set by its mass.
The mixing between wave fields can introduce a parametric structure, leading to an additional slow timescale.
In the narrow resonance regime or stable region near the boundary, this results in a slow envelope modulation accompanying the primary oscillation.
Such a two-timescale structure modifies the conventional single-frequency interpretation and can give rise to distinctive signatures across a wide range of modulation-based searches.

\begin{acknowledgments}
We thank Donghee Lee, Jiheon Lee, and Jaeok Yi for their helpful comments.
This work was supported in part by the National Research Foundation of Korea (Grant No. RS-2024-00352537).     
\end{acknowledgments}

\appendix

\section{Dynamics in dense environments}
\label{sec:dense}

In a substantially denser region, such as the galactic center, the dynamical effect of the mixing interaction becomes stronger.
The system may lie in the stable region or higher instability bands in the Mathieu parameter space. 
At sufficiently high densities, however, backreaction on both fields can no longer be neglected, and the Mathieu description becomes unreliable.

Here, we restrict our analysis to the homogeneous temporal dynamics induced by the density enhancement, rather than the spatial structure in the presence of the mixing interaction.
A consistent treatment of the spatial distribution would require solving the Schrödinger-Poisson equations~\cite{Hui:2021tkt}.
Related effects of mixing interactions in ultralight wave DM have been investigated~\cite{Guo:2020tla, Glennon:2023jsp, Mirasola:2024pmw}.

For an NFW-like Milky Way halo profile, the density enhancement can reach $\mathcal{O}(10^2-10^3)$ within the central $10-100\,\mathrm{pc}$~\cite{Navarro:1995iw, Navarro:1996gj, McMillan:2016jtx, Hooper:2016ggc}.
Assuming that the initial amplitudes of both components are rescaled by the same factor $\sqrt{D}$, 
the Mathieu parameters change as
\begin{align}
q(D) = D q_0,\; A(D) = A_0 + 2q_0(D-1),
\end{align}
where $D=1$ corresponds to the benchmark $(A_0, q_0)$ used in the main text.
As $D$ increases, the system moves along the trajectory in the $(A,q)$ parameter space of Fig.~\ref{AqPlotDen}.
 
\begin{figure}[tb]
    \centering
    \includegraphics[width=0.45\textwidth]{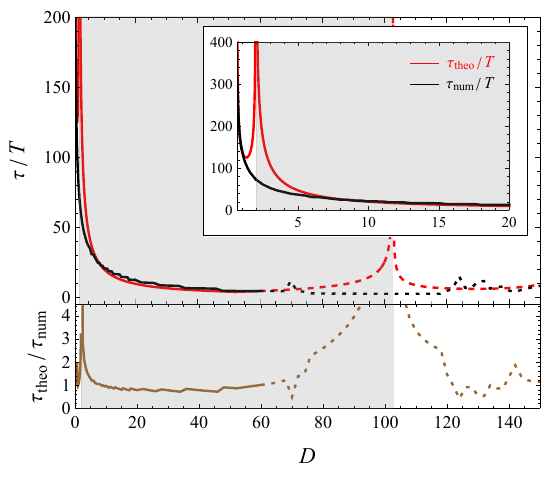}
    \caption{
    Comparison of the envelope timescale $\tau_{\rm num}/T$ from the full coupled numerical solution with parametric predictions $\tau_{\rm theo}/T$ along enhancement $D$.
    The white and gray regions correspond to the unstable and stable regions, respectively. 
    Near the boundary at $D\simeq \{0.67, \, 2, \, 102\}$, the prediction diverges, while the numerical result remains finite.
    The lower panel represents the ratio between theoretical and numerical estimations.
    For $D \gtrsim 60$, the Mathieu approximation becomes less valid, and $\tau_{\rm num}$ is no longer reliable due to the irregular oscillating pattern.
    }
    \label{timescale}
\end{figure}

Figure~\ref{timescale} compares the envelope timescale $\tau_{\rm num}/T$ extracted from the full coupled solution with the theoretical parametric estimate $\tau_{\rm theo}/T$ over the enhancement $D$.
In the numerical result, we define $\tau_{\rm num}$ from the separation between corresponding extrema according to the same cycle-counting convention adopted for $\tau_\mu$ and $\tau_\nu$.
Inside an instability band, the theoretical characteristic timescale is given by $\tau_\mu/T=1/\mu$. 
In a stable region, the prediction follows $\tau_\nu/T = 2/ |\nu-n|$, where $n$ is chosen as the nearest band index.
The numerical result captures the overall tendency of the theoretical prediction, while moderate deviations arise from nonlinear backreaction in the full coupled system.

The system leaves the first instability band at $D\simeq 0.67$ and $D \simeq 2$ and enters the adjacent stable region.
As the density is further increased, the trajectory approaches the second instability band around $D \simeq 102$. 
At these band boundaries, the parametric estimates formally predict a divergent envelope timescale. $\mu\rightarrow0$ when the boundary is approached from inside an instability band, while $|\nu-n|\rightarrow0$ when it is approached from the stable side.
The numerical evolution, however, retains a finite timescale.
Close to the band boundary, two modes become degenerate, so the second independent solution may acquire a secular form rather than describe finite-period beating.
In addition, the backreaction in the full nonlinear system makes parameters $(A,q)$ time-dependent, so the system does not remain at the fixed Mathieu boundary.
Consequently, the modulation period inferred from the parametric structure is not applicable in the immediate vicinity of either boundary.

As a representative example of the stable region, the $D=10$ evolution is shown in Fig.~\ref{AqPlotDen}.
It exhibits a regular wave-envelope behavior with $\nu\simeq 1.1$, corresponding to $\tau_\nu \simeq 20 T \simeq 3.2\,{\rm months}$.
As the trajectory moves away from the first-band boundary, the deviation $|\nu-n|$ increases and the corresponding timescale becomes shorter.
However, as the parameter approaches the second band, the tendency changes.

For $D \gtrsim 60$, we found that the Mathieu description becomes progressively less accurate since $M_\Phi$ is no longer approximately constant.
For $D=120$ in the plot, the parameter lies inside the second instability band.
The full numerical solution departs from the linear Mathieu prediction and shows an irregular envelope.
The extracted $\tau_{\rm num}$ does not represent a reliable envelope timescale in this regime.
Although the corresponding point lies in an instability region, nonlinear backreaction prevents indefinite exponential growth, satisfying the total energy conservation of the full two-field system.
The amplitudes of $\phi$ and $\Phi$ vary out of phase because of the energy exchange between the two fields.

\begin{figure}[t]
    \centering
    \includegraphics[width=0.52\textwidth]{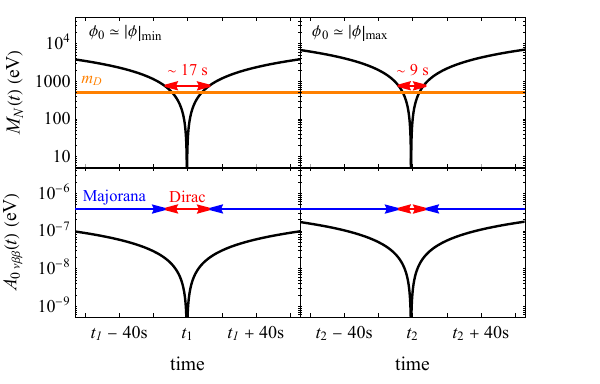}
    \caption{Time evolution of the Majorana mass $M_N(t)$, with the lower panel showing $\mathcal{A}_{0\nu\beta\beta}(t)$ around the Dirac regime ($m_D \gg M_N$), where the $0\nu\beta\beta$ decay is suppressed.
    When the DM gets the minimum amplitude $|\phi|_\mathrm{min}$ at $t_1$, the turn-off time interval is $\Delta T_\mathrm{Dirac} \simeq 17\,\mathrm{s}$.
    In contrast, at the maximum amplitude, $|\phi|_\mathrm{max}$ at $t_2$, the interval is shorter, $\Delta T_\mathrm{Dirac} \simeq 9\,\mathrm{s}$.
    The time difference $t_2-t_1 \simeq 0.47\,\mathrm{yrs}$ represents the slow oscillation.}
    \label{nuPlot}
\end{figure}

\section{Application to neutrino physics}
\label{appenNu}
As a phenomenological illustration of these more general phenomena, we focus on the neutrino sector and assume that the Majorana mass of the sterile neutrino $N$ is generated by the wave-envelope DM field $\phi$. 
\begin{align}
\mathcal{L} \supset - \frac12 g  \, \phi \, \overline{N^c} N + \mathrm{h.c.}
\label{Majorana}
\end{align}
The resulting Majorana mass is $M_N(t) = g \, \phi(t)$, where the negative sign can be absorbed by redefining the field.
The dynamical backreaction from the neutrino sector on the wave-envelope DM can be neglected, since the sterile-neutrino relic abundance is negligibly small in the current universe.

If the time-dependent Majorana mass $M_N(t)$ oscillates across the fixed Dirac mass term $m_D$, the neutrino can also oscillate between Dirac and Majorana types periodically depending on which mass term dominates~\cite{ChoeJo:2023ffp, Kim:2025xum}.
Especially, when the neutrino stays in the Dirac type ($m_D \gg M_N$), the lepton-number-violating process is strongly suppressed due to the smaller Majorana mass than the Dirac mass term~\cite{Doi:1982dn}.

A representative example of the lepton-number-violating process is the neutrinoless double beta ($0\nu\beta\beta$) decay~\cite{Schechter:1981bd}.
For the Dirac limit, the $0\nu\beta\beta$ decay process is effectively turned off.
The time interval $\Delta T_\mathrm{Dirac}$ of staying Dirac type for each turn-off is~\cite{ChoeJo:2023ffp}
\begin{align}
\Delta T_\mathrm{Dirac}  \simeq \frac{T}{\pi} \sin^{-1} \left( \frac{m_D}{g \phi_0} \right) ,
\end{align}
for $g\phi_0 > m_D$. 
Unlike wave DM with the static amplitude $\phi_0$, the wave-envelope DM inherits the modulated amplitude $\phi_0(t)$.
As a consequence, the turn-off time interval $\Delta T_\mathrm{Dirac}$ also varies. 
 
Figure~\ref{nuPlot} shows the magnified view around the Dirac transition.
The parameter choices are the same as in Fig.~\ref{phiPlot}, with additional inputs $g = 5 \times 10^{-11}$, and $m_D = 500\,\mathrm{eV}$, consistent with the observed active neutrino mass scale $m_\nu = 0.05\,\mathrm{eV}$ during the seesaw limit, assuming a single generation for simplicity.
We use this benchmark to illustrate the wave-envelope modulation in neutrino observables, without addressing the radiative stability.
The amplitude $\mathcal A_{0\nu\beta\beta}$ of the $0\nu\beta\beta$ decay, as a generalization of the effective electron neutrino mass $m_{ee}$, is evaluated using the nuclear matrix element to include the sterile neutrino contribution~\cite{Bolton:2019pcu}.

The turn-off occurs every $T/2 \simeq 2.4\,\text{days}$, but this can vary slightly due to the time-dependent DM mass.
For the small DM amplitude $\phi_0 \simeq |\phi|_\mathrm{min}$ at $t_1$, the turn-off time duration is $\Delta T_\mathrm{Dirac} \simeq 17\,\mathrm{s}$.
In contrast, when the DM amplitude is large $\phi_0 \simeq |\phi|_\mathrm{max}$ at $t_2$, the turn-off interval becomes relatively short as $\Delta T_\mathrm{Dirac} \simeq 9\,\mathrm{s}$.
The time difference $t_2 - t_1 \simeq \tau_\mu/4 \simeq 0.47\,\mathrm{yrs}$ between moments for the maximum and minimum DM amplitude is associated with the slow oscillation. 
This illustrates how the slow modulation induced by the parametric structure manifests in observable signals. 

The mass scale of the sterile neutrino reaches around $M_N \simeq 5\times 10^6\,\mathrm{eV}$ and the $0\nu\beta\beta$ decay amplitude becomes roughly $\mathcal A_{0\nu\beta\beta} \simeq 10^{-4}\,\mathrm{eV}$ for the Majorana case, but they are not shown in the magnified plot of Fig.~\ref{nuPlot}.
Since the DM amplitude was large in the early universe, the relevant constraints on the corresponding sterile neutrino mass scale~\cite{Bolton:2019pcu} can be avoided.
The neutrino mass variation can remain compatible with active neutrino flavor oscillation using the time-average analysis~\cite{Berlin:2016woy, Krnjaic:2017zlz, Dev:2022bae, Losada:2022uvr}.

\bibliography{ref}
 
\end{document}